\documentclass[preprint,showpacs,aps,floatfix]{revtex4} 
\usepackage{graphicx}
\include{epsf}
 
\begin{document}  

\title{Stochastic Simulations of Genetic Switch Systems} 

\author{Adiel Loinger$^1$, Azi Lipshtat$^2$, 
Nathalie Q. Balaban$^1$ and Ofer Biham$^1$
}  
\affiliation{  
$^1$ Racah Institute of Physics,   
The Hebrew University,   
Jerusalem 91904,   
Israel
\\
$^2$ Department of Pharmacology and Biological Chemistry,
Mount Sinai \\ School of Medicine,
New York, NY 10029, USA
}  
 
\begin{abstract}  

Genetic switch systems with mutual repression of two
transcription factors are studied using deterministic 
methods (rate equations)
and 
stochastic methods 
(the master equation
and Monte Carlo simulations).
These systems exhibit bistability, namely
two stable states such that 
spontaneous transitions between them are rare. 
Induced transitions may take place as a
result of an external stimulus.
We study several variants of the genetic
switch and examine the effects
of cooperative binding, exclusive binding, protein-protein
interactions and degradation of bound repressors.
We identify the range of parameters
in which bistability takes place, enabling the system
to function as a switch. 
Numerous studies have concluded 
that cooperative binding is 
a necessary condition for the emergence of bistability
in these systems.
We show that a suitable combination of network structure and
stochastic effects gives rise to bistability even
without cooperative binding. 
The average time between spontaneous transitions is evaluated
as a function of the biological parameters.

\end{abstract}

 
\maketitle  

\section{Introduction}
\label{Introduction}

Recent advances in quantitative measurements 
of gene expression at the single-cell level 
\cite{Elowitz2002,Ozbudak2002}
have brought new insight on the importance
of stochastic fluctuations 
in genetic circuits
\cite{Mcadams1997}.
The role of fluctuations is enhanced due to 
the discrete nature of the transcription factors
and their binding sites, which may appear in low copy numbers
\cite{Becskei2000,Kaern2005}.
As a result,
populations of genetically identical cells may 
show significant variability.
Stochastic behavior
may invoke oscillations
\cite{Vilar2002}
and spatio-temporal patterns
\cite{Shnerb2000},
which are unaccounted for by
macroscopic chemical rate equations. 
Genetic circuits with
feedback mechanisms may exhibit bistability,
namely, two distinct stable states which
can be switched by an external 
signal
\cite{Atkinson2003}.
A low rate of spontaneous switching events
may also take place.
To qualify as a switch, this rate
must be much lower than the rates of the relevant
processes in the cell, namely transcription, translation,
and degradation of transcription factors.
Genetic switches, such as the 
phage $\lambda$ switch, 
give rise to different cell fates
\cite{Ptashne1992}.
In this switch, $\lambda$
phages infect {\it E. coli} bacteria and 
can exist in two exclusive states,
one called 
lysogeny
and the other called 
lysis. 
When the phage enters its host, it integrates itself into the
host's DNA and is duplicated by cell division.
It codes for proteins that can identify stress in the host cell.
In case of stress, the phage transforms into the lysis state.
In this state, it kills the host cell, using its DNA to produce
many copies of the phage, which are released and later infect other
cells.
Other switch circuits exist in the metabolic systems of cells.
These switches determine which type of sugar the cell will digest
\cite{Lewis2005}.
The genetic switch may also serve as a memory unit of the cell,
and help determine its fate during cell differentiation.

Recent advances enable the construction of genetic circuits
with desired properties, that are determined by the network
architecture.
These networks are constructed from available components, namely 
genes and promoters. They do not require the manipulation of the 
structure of proteins and other regulatory elements at the
molecular level.
These genes and promoters are often inserted into plasmids
rather than on the chromosome.
A synthetic toggle switch, that consists of two repressible promoters
with mutual negative regulation,
was constructed in {\it E. coli} and the conditions for bistability
were examined
\cite{Gardner2000}.
The switching between its two states was demonstrated using 
chemical and thermal induction.
More recently, such circuit was found to exist in a natural
system in which two mutual repressors regulate
the differentiation of myeloid progenitors
into either macrophages or neutrophils
\cite{Laslo2006}.

In this paper we analyze   
the genetic toggle switch using deterministic and stochastic methods.
In this simple genetic circuit, 
two proteins, $A$ and $B$,
negatively regulate each other's synthesis.
The regulation is performed at the transcription level, namely
the production of protein $A$ is 
negatively regulated by protein $B$,
through binding of $n$ copies of $B$
to the $A$ promoter (and vice versa). 
This process can be
modeled by a Hill function, which 
reduces the production rate of
$A$ 
by a factor of
$1+k[B]^n$,
where 
$[B]$
is the concentration of $B$ proteins
in the cell,
$k$ is a parameter and 
$n$ is the Hill coefficient
\cite{Hill}.
In case that $n=1$,
the binding of a single protein is 
sufficient in order to perform the negative regulation, while
for $n>1$ the {\it cooperative binding} of two or more
proteins is required.

One may expect this circuit to function as a switch,
with two stable states, one dominated by $A$ proteins
and the other dominated by $B$ proteins.
When the population of $A$ proteins is larger than the population of $B$ 
proteins, the $A$ proteins suppress the production of $B$ 
proteins. Under these conditions, 
the production of $A$ proteins will not be suppressed
much by the small $B$ population. 
Therefore, the system approaches a state
rich in $A$ proteins and poor in $B$ proteins. 
Similarly, the system may approach a state rich 
in $B$ proteins and poor
in $A$ proteins.

To qualify as a switch, the system should be 
bistable. 
In the deterministic description, bistability is defined
as the existence of two stable steady state solutions 
of the rate equations.
This description does not account for the possibility of
spontaneous transitions between the two states.
In the stochastic description, spontaneous transitions
do take place. Therefore, the condition for bistability
is that the rate of
spontaneous switching events 
(due to random fluctuations rather than 
an external signal) 
is much lower than the rates of
all other relevant processes
in the system. 

Rate equations provide the average concentrations of $A$ and $B$
proteins in a population of cells.
In these equations, bistability emerges at a bifurcation
point, where two stable states emerge.
Rate equations do not include fluctuations and do
not account for the possibility of spontaneous transitions
between the two states.
The master equation provides the probability distribution
of the populations of $A$ and $B$ proteins.
The two bistable states appear as two distinct peaks
in this distribution.
Monte Carlo simulations enable to follow the fluctuations
in a single cell and to evaluate the rate of spontaneous
switching events.

We examine the conditions for the system to become 
a switch, and
calculate the rate of spontaneous transitions between its
two states.
This is done for several variants of the toggle switch.
In particular, 
we focus on switch systems in which the repression in done
without cooperative binding (namely, $n=1$). 
Numerous studies have concluded,
using rate equations,
that cooperative binding is
a necessary condition for the emergence of bistability
\cite{Gardner2000,Cherry2000,Warren2004,Warren2005,Walczak2005}. 
Below we show,
using a combination of deterministic and stochastic simulation methods,
that this is not the 
case, namely a bistable switch can exist even in
the absence of cooperative binding.
In particular, we show that bound-repressor degradation (BRD)
and protein-protein interactions (PPI)
give rise to bistability,
without cooperative binding,
even at the level of rate equations.
These results are confirmed by stochastic simulations using
the master equation and Monte Carlo methods.
We also consider the exclusive switch,
in which the $A$ and $B$ repressors cannot be bound simultaneously
due to overlap between their promoter sites.
This system
exhibits bistability only when 
stochastic fluctuations are taken into account.
The rate of spontaneous transitions between the two states
is calculated as a function of the biological parameters.

The paper is organized as follows.
In Sec.
\ref{sec:general}
we consider the basic version called the general switch.
Several variants of this circuit are considered in the
Sections that follow.
The exclusive switch is studied in
Sec.
\ref{sec:exclusive},
the BRD switch is considered in 
Sec.
\ref{sec:BRD}
and the PPI switch is analyzed in 
Sec.
\ref{sec:PPI}.
The effects of cooperative binding are studied in
Sec.
\ref{sec:cooperative}.
The response of toggle switch systems to external 
signals is examined in 
Sec.
\ref{sec:response}.
The results are discussed in 
Sec.
\ref{sec:discussion}
and summarized in
Sec. 
\ref{sec:summary}.

\section{The General Switch (Without Cooperative Binding)}
\label{sec:general}

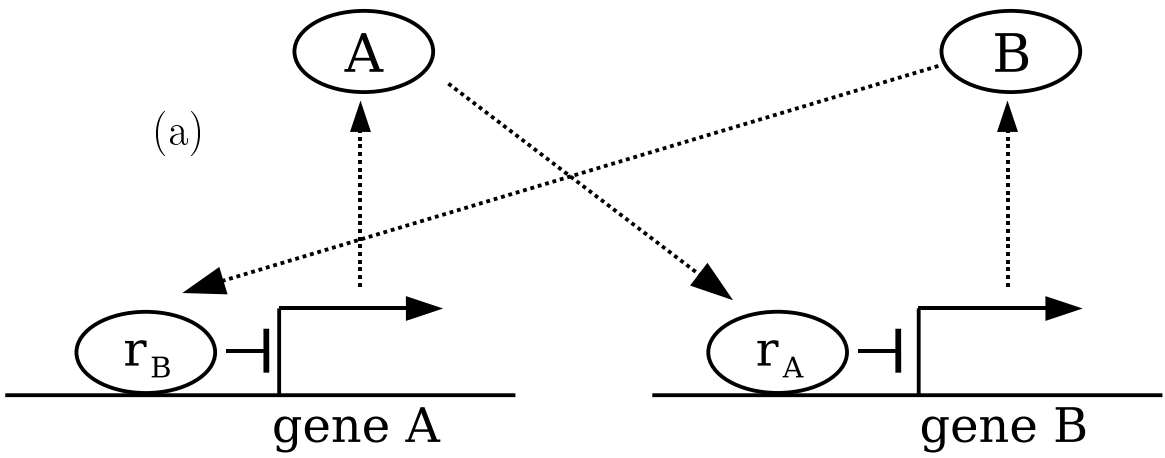
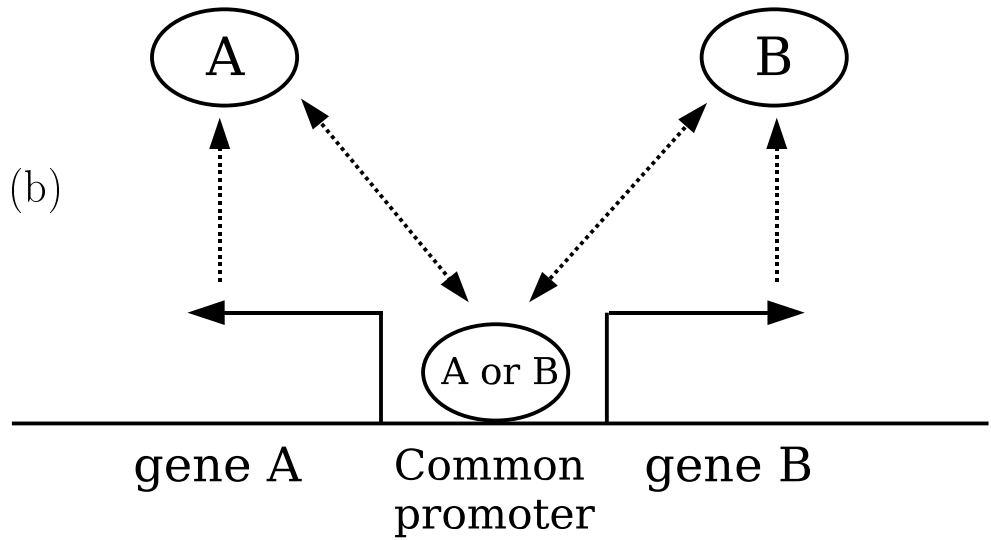
\begin{figure}
\caption{Schematic illustrations of (a) the general switch circuit,
that includes two transcription factors, $A$ and $B$, 
which negatively regulate each other's synthesis; 
(b) the exclusive switch,  
in which there is an overlap between the 
promoter sites of $A$ and $B$ proteins, 
so they cannot be bound simultaneously.
}
\label{fig1}
\end{figure}

The general switch consists of two transcription factors,
$A$ and $B$, that negatively regulate each other's
synthesis
\cite{Warren2004,Warren2005}.
A schematic description of 
this circuit is given in
Fig. \ref{fig1}(a).
The regulation is done by the binding of a protein to
the promoter site of the other gene, blocking the access
of the RNA polymerase and suppressing the transcription 
process.
In this circuit there is no cooperative binding, namely
the regulation process is performed by a single bound
protein.

The concentrations of free $A$ and $B$ proteins in the cell
are denoted by $[A]$ and $[B]$, respectively (by concentration
we mean the average copy number of proteins per cell).
The copy numbers of the bound proteins, are denoted by 
$[r_A]$ and $[r_B]$,
where $r_A$ is a bound $A$ protein that monitors the production of $B$, 
while $r_B$ is a bound $B$ protein that monitors the production of $A$.
Note that there is at most one bound repressor of each type at any
given time, and thus
$0 \le r_A,r_B \le 1$.
For simplicity, we ignore the mRNA level and combine the 
processes of transcription and translation as a single 
step of synthesis
\cite{mRNA}. 

The maximal production rate of protein $X$ is denoted by
$g_X$ (s$^{-1}$), $X=A,B$.
The degradation rate of protein $X$ is given by
$d_X$ (s$^{-1}$).
While the structure of the circuits studied here
is symmetric,
the rate constants can be different for $A$ and $B$. 
However, for simplicity we use symmetric parameters,
i.e. $g=g_A=g_B$ and $d=d_A=d_B$.
The binding rate of proteins to the promoter
is denoted by $\alpha_0$ (s$^{-1}$)
and the dissociation rate
by $\alpha_1$ (s$^{-1}$). 

\subsection{Rate Equations}

The dynamics of the general switch circuit 
is described by the rate equations
\cite{Lipshtat2005,Lipshtat2006}

\begin{eqnarray}
\dot{[A]} &=& g_A (1-[r_B]) - d_A [A]-
\alpha_0 [A] \left (1-[r_A] \right) + \alpha_1[r_A]  \nonumber \\
\dot{[B]} &=& g_B(1-[r_A])-d_B[B]-
\alpha_0[B]\left(1-[r_B]\right)+\alpha_1[r_B] \nonumber \\
\dot{[r_A]} &=& \alpha_0[A]\left(1-[r_A]\right)-\alpha_1[r_A] \nonumber \\
\dot{[r_B]} &=& \alpha_0[B]\left(1-[r_B]\right)-\alpha_1[r_B].
\label{eq:general_switch_rate}
\end{eqnarray}

\noindent
It is commonly assumed that the 
binding-unbinding processes 
are much faster than other 
processes in the circuit, namely 
$\alpha_0,\alpha_1 \gg d_X,g_X$. 
This means that the relaxation times of $[r_X]$ 
are much shorter than other relaxation 
times in the circuit. 
Under this assumption,
one can take the 
time derivatives of $[r_X]$
to zero, even if the system is away from
steady state.
This brings the rate equations to the 
standard Michaelis-Menten form

\begin{eqnarray}    
\dot{[A]} &=& \frac{g}{1+k[B]} - d [A] \nonumber \\
\dot{[B]} &=& \frac{g}{1+k[A]} - d [B],
\label{eq:general_switch_michaelis_menten}
\end{eqnarray}    

\noindent
where symmetric parameters are used, and
$k=\alpha_0/\alpha_1$ 
is the repression strength.
For a given population of free $X$ repressors,
the parameter $k$  controls the value of $[r_X]$.
The limit of weak repression,
$[r_X] \ll 1$,
is obtained when
$k[X] \ll 1$,
while the limit of strong repression,
$[r_X] \simeq 1$,
is obtained for
$k[X] \gg 1$. 

The meaning of bistability at the level of rate equations is
that at steady state the equations exhibit two distinct positive
solutions.
In this particular class of circuits, 
one solution is dominated by $A$ proteins and the other
is dominated by $B$ proteins.
Starting from any initial state, the system will converge
to one of these solutions.
The solutions are stable, so the possibility of spontaneous 
transitions, induced by stochastic fluctuations, 
is not included in the rate equation description.

The steady state solutions of Eqs.
(\ref{eq:general_switch_rate})
and 
(\ref{eq:general_switch_michaelis_menten})
are identical.
We will now show that
these equations have only one positive steady-state solution. 
To this end, we first
take $\dot{[A]}=\dot{[B]}=0$ 
in Eq.
(\ref{eq:general_switch_michaelis_menten}).
We multiply each equation by the denominator
of the Hill function that appears in it. 
We obtain:

\begin{eqnarray}    
g - d[A] - kd[A][B] &=& 0 \nonumber \\
g - d[B] - kd[A][B] &=& 0.
\label{eq:general_switch_hill1}
\end{eqnarray}    

\noindent
Subtracting one equation from the other we get 
$d([A]-[B])=0$ 
and therefore 
$[A]$ must be equal to $[B]$ at steady state.
The steady state values of 
$[A]$ and $[B]$ can be easily found. 
Inserting
$[A]=[B]$ into 
Eq.~(\ref{eq:general_switch_hill1}) 
we obtain a quadratic equation
whose only positive solution is 

\begin{equation}
[A] = [B] =  \frac {-1+\sqrt{1+4kg/d}}{2k}.
\end{equation}

\noindent
Standard linear stability analysis shows that 
this solution is always stable.
    
As a result, we conclude that at the level of rate equations
the general switch, without cooperative binding,
does not exhibit bistability.
In Sec. 
\ref{sec:cooperative}
we consider the case of cooperative binding, where
the rate equations do exhibit bistability.

\subsection{Master Equation}

In order to account for stochastic effects  
and to obtain insight on 
the reason that this system is not bistable, 
the master equation approach is applied
\cite{Mcadams1997,Arkin1998,Kepler2001,Paulsson2000,Paulsson2004}.
In this case, we consider
the probability distribution function
$P(N_A,N_B,r_A,r_B)$.
It is the probability for a cell to 
include $N_X$ copies of free protein $X$
and  $r_X$ copies of the bound $X$ repressor,
where $N_X=0,1,2,\dots$,  
and $r_X = 0,1$.
The master equation for the general switch takes the form

\begin{eqnarray}
&& \dot{P}(N_A,N_B,r_A,r_B) = 
g_{\rm A} \delta_{r_B,0} 
[P(N_A-1,N_B,r_A,r_B) - P(N_A,N_B,r_A,r_B)]  \nonumber\\
&& + g_{\rm B} \delta_{r_A,0} 
[P(N_A,N_B-1,r_A,r_B) - P(N_A,N_B,r_A,r_B)] \nonumber\\
&& + d_{\rm A} [(N_A+1) P(N_A+1,N_B,r_A,r_B) - N_A P(N_A,N_B,r_A,r_B)]  
\nonumber\\
&& + d_{\rm B} [(N_B+1) P(N_A,N_B+1,r_A,r_B) - N_B P(N_A,N_B,r_A,r_B)]  
\nonumber\\
&& + \alpha_0 [(N_A+1) \delta_{r_A,1} P(N_A+1,N_B,r_A-1,r_B) - 
              N_A \delta_{r_A,0} P(N_A,N_B,r_A,r_B)]  \nonumber\\
&& + \alpha_0 [(N_B+1) \delta_{r_B,1} P(N_A,N_B+1,r_A,r_B-1) - 
              N_B \delta_{r_B,0} P(N_A,N_B,r_A,r_B)]  \nonumber\\
&& + \alpha_1 [\delta_{r_A,0} P(N_A-1,N_B,r_A+1,r_B) -
             \delta_{r_A,1} P(N_A,N_B,r_A,r_B)]  \nonumber\\
&& + \alpha_1 [\delta_{r_B,0} P(N_A,N_B-1,r_A,r_B+1) -
             \delta_{r_B,1} P(N_A,N_B,r_A,r_B)], 
\label{eq:general_switch_master}
\end{eqnarray}

\noindent
where 
$\delta_{i,j}=1$ for $i=j$ and $0$ otherwise.
The $g_X$ terms account for the production of proteins.
The $d_X$ terms account for the degradation of free proteins,
while the $\alpha_0$ ($\alpha_1$) terms describe the 
binding (unbinding) of proteins to (from) the promoter site. 
In numerical integration, the master equation must be truncated
in order to keep the number of equations finite. 
This is done by setting suitable upper cutoffs,
$N_{\rm A}^{\rm max}$ and $N_{\rm B}^{\rm max}$,
on the populations sizes of free proteins.
In order to maintain the accuracy of the calculations,
the probability of population sizes beyond the cutoffs
must be sufficiently small.

The master equation has a single steady state solution, 
which is always stable 
\cite{VanKampen1992}.
The criterion for bistability  
is that the steady state solution 
$P(N_A,N_B,r_A,r_B)$ 
exhibits two distinct regions (peaks) of high  
probabilities, 
separated by a gap in which the probabilities are very small.
These two regions correspond to the two  
states in which the system is likely to be.
If the transition rate between the peaks
is small enough, 
the system is indeed a bistable switch.
Note, that in this case, averages of the form
  
\begin{equation}
\langle N_{\rm X} \rangle =
\sum_{N_{\rm A}=0}^{N_{\rm A}^{\rm max}} 
\sum_{N_{\rm B}=0}^{N_{\rm B}^{\rm max}} 
\sum_{r_{\rm A}=0}^1 
\sum_{r_{\rm B}=0}^1 
N_{\rm X} P(N_A,N_B,r_A,r_B),
\label{eq:averagex}
\end{equation}

\noindent
where $X=A$, $B$,
do not reflect the complex structure of the 
probability distribution.
These can be considered as averages over many
cells, some dominated by $A$ and others dominated by $B$
proteins, such that the total populations of the two species
are about the same.

To examine  
the existence of bistability we consider
the marginal probability distribution 

\begin{equation}
P(N_A,N_B) = 
\sum_{r_{\rm A}=0}^1
\sum_{r_{\rm B}=0}^1
P(N_A,N_B,r_A,r_B).
\end{equation}
\noindent

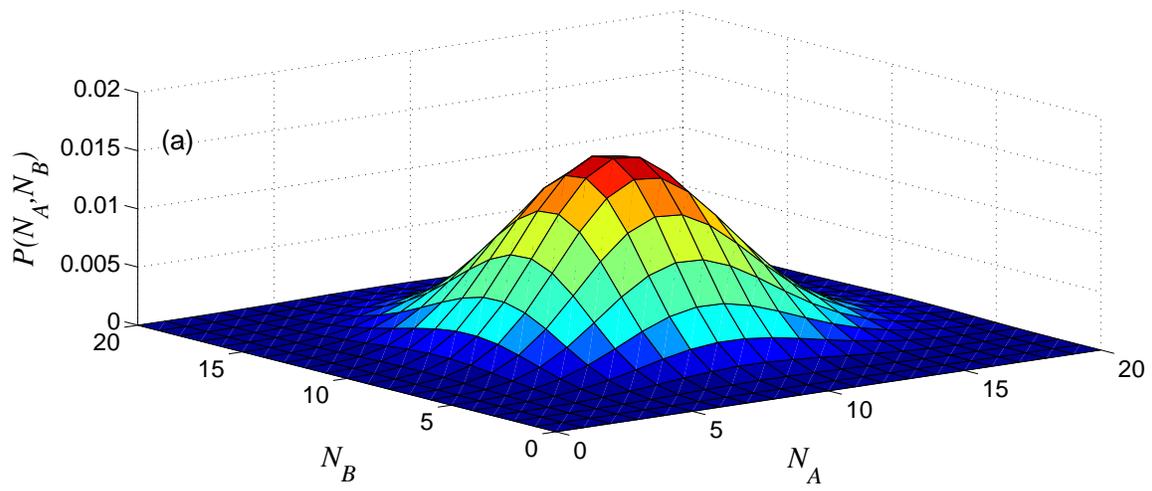
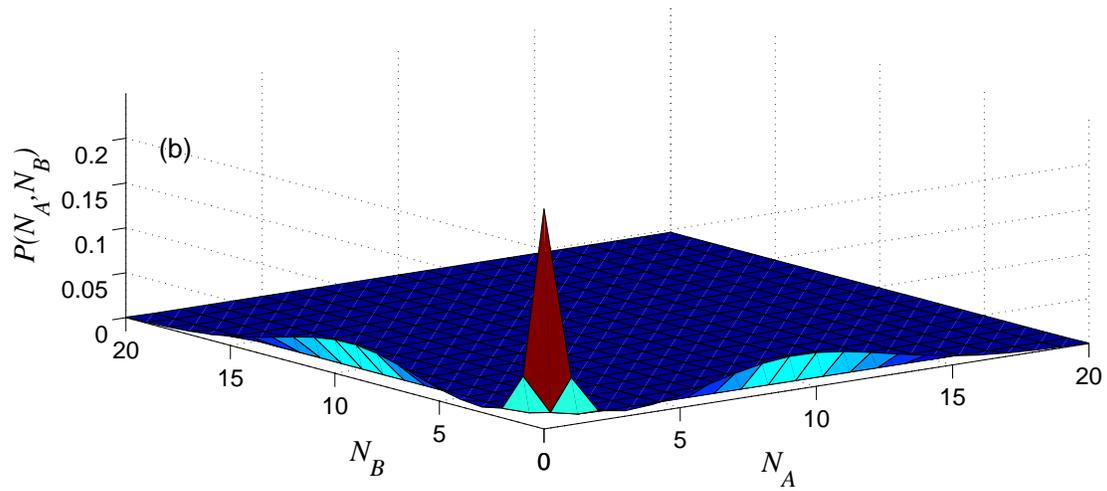
\begin{figure}
\caption{
(Color online)
The probability distribution
$P(N_A,N_B)$ 
for the general switch, 
under conditions of 
(a) weak repression ($k=0.005$) 
where there is one symmetric peak;
and 
(b) strong repression ($k=50$)
where three peaks appear, one dominated by $A$, the
second dominated by $B$ and the third in which 
both species are mutually suppressed. 
The weights of the three peaks are about the same.
}
\label{fig2}
\end{figure}

\noindent
This probability distribution was calculated 
for a broad range of parameters.
Two representative examples are shown in 
Fig. \ref{fig2}.

Under conditions of weak repression (small k),
$P(N_A,N_B)$ 
exhibits a single peak for which 
$N_A \approx N_B \approx g/d$
[Fig. \ref{fig2}(a)],
in agreement with the rate equations.
This is due to the fact that the
repression is weak, and the 
$A$ and $B$ populations are almost uncorrelated.
In this case, the cell will contain roughly the same
amount of $A$ and $B$ proteins. 

For strong repression, the distribution $P(N_A,N_B)$ 
exhibits a peak dominated by $A$ proteins
and a peak dominated by $B$ proteins,	
as expected for a bistable system. 
However, a third peak appears near the origin,
in which both populations of free proteins 
are suppressed
[Fig. \ref{fig2}(b)]. 
This peak represents a dead-lock situation, caused by the 
fact that both $A$ and $B$ repressors can be bound
simultaneously, each bringing to a halt the production of the other specie.
The third peak provides a corridor through which the probability can
flow between the other two peaks.
As a result, the system can quickly switch between the
$A$-dominated and the $B$-dominated states.

In addition to the solution of the master equation, 
Monte Carlo simulations
have been performed. 
In these simulations one can follow the time evolution
of the populations of free and bound proteins
in a single cell. 
In 
Fig. \ref{fig3}(a)
we present the population sizes of free proteins vs. time
for the general switch.
It is clear that the cell can indeed
be in one of three states: a state rich in $A$, 
a state rich in $B$ and a state
in which both proteins are in very low copy numbers. 
We conclude that a necessary condition for the system to become a switch
is to prevent this dead-lock situation in which both protein
populations are suppressed simultaneously. 
Below we present several 
variants of the circuit in which the third peak is suppressed, 
giving rise to a bistable switch.

\section{The Exclusive Switch}
\label{sec:exclusive}

The first variant we consider is the exclusive switch,
depicted in 
Fig.~\ref{fig1}(b).
In this circuit 
there is an overlap between the promoters 
of $A$ and $B$.
As a result, there is
no room for both  
$A$ and $B$ proteins to be bound
simultaneously.
Exclusive binding is encountered in nature,  
for example, in the 
lysis-lysogeny switch of phage $\lambda$
\cite{Ptashne1992}.

It was shown that 
in presence of cooperative binding,
the exclusive switch is more stable than
the general switch
\cite{Warren2004,Warren2005}.
This is because in the exclusive switch the access of the
minority proteins to the promoter site is blocked by the
dominant proteins.
Here we show that in the exclusive switch, 
stochastic effects give rise to bistability even
without cooperativity between the transcription factors.
The dead-lock situation in prevented in this case, since 
$A$ and $B$
repressors cannot be bound simultaneously. 

\subsection{Rate Equations}

To model 
the exclusive switch,
recall that the variable
$[r_A]$  ($[r_B]$)  
is actually the fraction of 
time in which the promoter 
is occupied by a bound $A$ ($B$) protein
\cite{Lipshtat2005}.
The fraction of time in which the promoter is vacant is
thus
$1 - [r_A] - [r_B]$.
Incorporating this into 
Eq.~(\ref{eq:general_switch_rate})
gives rise to the following modification:
in the $\alpha_0$ terms,
each appearance of $[r_A]$ or $[r_B]$
should be replaced by $[r_A]+[r_B]$.
With this modification, the rate equations take
the form

\begin{eqnarray}
\dot{[A]} &=& g (1-[r_B]) - d [A]-
\alpha_0 [A] \left (1-[r_A]-[r_B] \right) 
+ \alpha_1 [r_A]  \nonumber \\
\dot{[B]} &=& g (1-[r_A]) - d [B] -
\alpha_0 [B]\left(1-[r_A]-[r_B] \right)
+\alpha_1 [r_B] \nonumber \\
\dot{[r_A]} &=& \alpha_0 [A] 
\left(1- [r_A] - [r_B] \right)-\alpha_1 [r_A]  \nonumber \\
\dot{[r_B]} &=& \alpha_0 [B] 
\left(1- [r_A] - [r_B] \right)-\alpha_1 [r_B].
\label{eq:exclusive_rate}
\end{eqnarray}

\noindent
Under steady state conditions, the rate equations can be reduced
to the Michaelis-Menten form

\begin{eqnarray}
\dot{[A]} &=& \frac{g}{1+{k[B]/(1+k[A])}} - d[A] \nonumber \\
\dot{[B]} &=& \frac{g}{1+{k[A]/(1+k[B])}} - d[B],
\label{eq:exclusive_michaelis_menten}
\end{eqnarray}   

\noindent
where, as before, $k=\alpha_0/\alpha_1$.
We will now show that even for the case of the 
exclusive switch, the rate equations still exhibit 
a single solution, thus there is no bistability.
This is done by 
taking $\dot{[A]}=\dot{[B]}=0$ and getting rid of the
denominators, by repeated multiplications.
The resulting equations are

\begin{eqnarray}
g + (kg- d) [A]  -kd[A] ([A]+[B]) &=& 0 \nonumber \\
g + (kg- d) [B]  -kd[B] ([A]+[B]) &=& 0. 
\label{Brep2}
\end{eqnarray}   

\noindent
By subtraction of one equation from the other, 
we find that

\begin{equation}
\{kg-d - kd ([A]+[B])\}([A]-[B]) = 0. 
\label{Brep3}
\end{equation}   

\noindent
The positive, symmetric solution, $[A]=[B]$, is given by

\begin{equation}
[A] = \frac{(kg-d) + \sqrt{(kg+d)^2 +4kgd}}{4kd}.
\end{equation}   

\noindent
The other, non-symmetric solution, given by

\begin{equation}
kg-d -kd ([A]+[B]) = 0
\label{eq:nonsymmsol}
\end{equation}

\noindent
is inconsistent with
Eq.~(\ref{Brep2})
unless $g=0$, 
namely there is no production of $A$ and $B$ proteins,
which immediately leads to $[A]=[B]=0$.
Under these conditions the solution of Eq.
(\ref{eq:nonsymmsol})  
is 
$[A]+[B]=-1/k$, which requires a negative population size
and thus makes no physical sense.
Therefore, the only solution for $g>0$
is the symmetric solution, $[A]=[B]$.
Thus, the rate equations do not support a bistable
solution for the exclusive switch for any choice
of the parameters.

\subsection{Master Equation}

To account for the effects of fluctuations, 
we now describe the exclusive switch using 
the master equation.
It is similar to to master equation 
for the general switch given by 
Eq.~(\ref{eq:general_switch_master}),
except for the following modifications:
(a) In the $\alpha_0$ and $\alpha_1$ terms,
each time $\delta_{r_A,j}$ ($\delta_{r_B,j}$),
$j=0,1$,
appears it should be multiplied by
$\delta_{r_B,0}$, ($\delta_{r_A,0}$);
(b) The constraint 
$P(N_A,N_B,1,1)=0$ should be imposed.
Implementing these changes we obtain the following equation:

\begin{eqnarray}
\label{eq:exclusive_switch_master}
&& \dot{P}(N_A,N_B,r_A,r_B) = 
   g_{\rm A} \delta_{r_B,0} 
[P(N_A-1,N_B,r_A,r_B) - P(N_A,N_B,r_A,r_B)]  \nonumber\\
&& + g_{\rm B} \delta_{r_A,0} 
[P(N_A,N_B-1,r_A,r_B) - P(N_A,N_B,r_A,r_B)] \nonumber\\
&& + d_{\rm A} [(N_A+1) P(N_A+1,N_B,r_A,r_B) 
- N_A P(N_A,N_B,r_A,r_B)]  \nonumber\\
&& + d_{\rm B} [(N_B+1) P(N_A,N_B+1,r_A,r_B) 
- N_B P(N_A,N_B,r_A,r_B)]  \nonumber\\
&& + \alpha_0 \delta_{r_B,0} 
[(N_A+1) 
\delta_{r_A,1}
P(N_A+1,N_B,r_A-1,r_B) - 
N_A \delta_{r_A,0}
P(N_A,N_B,r_A,r_B)] 
 \nonumber\\
&& + \alpha_0 \delta_{r_A,0} 
[(N_B+1) \delta_{r_B,1} P(N_A,N_B+1,r_A,r_B-1) - 
              N_B \delta_{r_B,0} P(N_A,N_B,r_A,r_B)] 
 \nonumber\\
&& + \alpha_1 [\delta_{r_A,0} P(N_A-1,N_B,r_A+1,r_B) -
             \delta_{r_A,1} P(N_A,N_B,r_A,r_B)]  \nonumber\\
&& + \alpha_1 [\delta_{r_B,0} P(N_A,N_B-1,r_A,r_B+1) -
             \delta_{r_B,1} P(N_A,N_B,r_A,r_B)]. 
\end{eqnarray}

\noindent
For the exclusive switch, as for the general switch,
under conditions of weak repression,
$P(N_A,N_B)$ exhibits a single peak
[Fig. \ref{fig4}(a)]
that satisfies $N_A \approx N_B \approx g/d$.
However, as the repression strength increases 
two distinct peaks begin to form.
For intermediate values of $k$ these peaks are still connected, 
by a corridor of non-vanishing probabilities
[Fig. \ref{fig4}(b)]. 
Monte Carlo simulations show that for intermediate values of $k$,
the system indeed exhibits two states, 
one rich in $A$ and the other rich in $B$, 
but rapid transitions occur between 
them.

For strong repression,
the distribution 
$P(N_A,N_B)$ 
exhibits two peaks which are separated by 
a region with vanishing probabilities
[Fig. \ref{fig4}(c)].
In one peak the $A$ population is suppressed, while in the 
other peak the $B$ population is suppressed, 
as expected for a bistable system. 
The average population of the dominant protein specie in each peak is 
$\langle N_{\rm X} \rangle \approx g/d$,
while the population of the suppressed specie is 
$\langle N_{\rm X} \rangle \approx 0$. 
Monte Carlo simulations show that in this case the average
time between spontaneous transitions is much longer. 
The typical switching time  for the case shown
in Fig. \ref{fig3}(b)
is around $10^{5}$ seconds. 
It is much longer than the time-scales of the transcription,
translation and degradation processes.
It is also longer than the time between cell divisions
which is of the order of $10^3-10^4$ seconds.
The Monte Carlo results clearly
show a large number of failed attempts in which a protein of the minority 
specie binds to the promoter and then unbinds again, without causing 
the system to flip.

\subsection{Analysis of Switching Times}

To evaluate the switching times
we performed the following procedure.
We initialized the master equation in a state which
is completely dominated by A proteins, 
namely,
$P(N_A=\lfloor g/d \rfloor,N_B=0,r_A=0,r_B=0)=1$
(where $\lfloor \  \rfloor$ represents the integer part),
and all other probabilities vanish.
The master equation was then integrated numerically 
and $P(N_A,N_B)$ was calculated 
as a function of time. 
The function 
$f(t)=P(N_A>N_B)-P(N_A<N_B)$
was found to decay exponentially 
from its initial value,
$f(0)=1$, to zero,
according to
$f(t)=\exp(-t/\tau)$. 
The time constant
$\tau$ is defined as the switching time
\cite{TauMEMC}. 
Its inverse, $\tau^{-1}$, 
is referred to as the switching rate. 

Using this procedure, we examined the dependence of the
switching time, $\tau$ on the protein synthesis rate, $g$
[Fig. \ref{fig5}(a)], 
the degradation rate, $d$
[Fig. \ref{fig5}(b)], 
and the repression strength, $k$ 
[Fig. \ref{fig5}(c)].
In the parameter range in which bistability takes place,
we obtain that
(a) $\tau \sim g$,
(b) $\tau \sim 1/d^2$
and
(c) $\tau \sim k$.
Concerning 
Fig. \ref{fig5}(c),
note that system exhibits bistability only in the regime in which $k$
is large
\cite{OnlyK2006}. 
For $k<1$ , $\tau \sim 100-1000$ (s), which is the typical time-scale
of other processes in the cell.
Only for $k \gtrsim 10$, 
$\tau$ becomes significantly larger than 
the time scales of other processes, and 
the system can function as a stable switch. 
The scaling properties of the switching time can be summarized by

\begin{equation}
\tau \sim {\alpha_0 \over \alpha_1} {g \over d^2}.
\label{eq:exclusive_scaling}
\end{equation}

\noindent
This result can be reproduced by a simple argument.
Consider an initial state in which the system is dominated by
$A$ proteins, while the population of $B$ proteins is suppressed,
namely
$[A] \gg [B]$.
In this situation the promoter site is occupied by an $A$ protein
during most of the time. 
In order that the switch will flip, 
the bound A protein must
unbind (at rate $\alpha_1$). 
Then, a $B$ protein (rather than an $A$ protein)
should bind to the promoter.
The probability for this to happen is
$\sim [B]/[A]$. 
This $B$ protein
should remain bound long enough in order to build up a sufficiently
large population of $B$ proteins.
On average, the $B$ protein stays bound $1/\alpha_1$ (s), 
during which $g/\alpha_1$ proteins of type $B$ are produced.
After the $B$ repressor will unbind, the probability that 
the next protein that binds will be of type $B$ rather than $A$,
is thus
$\sim (g/\alpha_1)/[A]$ 
(neglecting the degradation of $A$ proteins,
because $\alpha_1 \gg d$).
Following this argument, 
the switching rate is given by

\begin{equation}
\tau^{-1} \sim
\alpha_1 
\times {[B] \over [A]} 
\times { {g} \over {\alpha_1 [A]} } 
= g { [B] \over [A]^2 }.
\label{eq:exclusive_scaling2}
\end{equation}

\noindent
From the Michaelis-Menten equations we obtain that 

\begin{equation}
\frac{[B]}{[A]} = \frac{1}{1 + k [A]} \approx \frac{1}{k[A]}, 
\end{equation}

\noindent
since 
for strong repression
$k[A] \gg 1$. 
Inserting this result into Eq. 
(\ref{eq:exclusive_scaling2})
and
using
and 
$[A] \approx g/d$
we find that

\begin{equation}
\tau = \frac{kg}{d^2}.
\end{equation}

\noindent
This result can be considered as the leading term in the expansion
of $\tau$ in powers of $g$, $d$ and $k$. 
This leading term turns out to provide a
very good approximation to simulation results. 
For example
for $g=0.2$,
$d=0.005$,
$\alpha_0=0.2$,
and
$\alpha_1=0.01$
(s$^{-1}$) 
we get 
$\tau=1.6 \cdot 10^5$ (s),
which agrees perfectly with 
the results of Monte Carlo simulations. 

From Eq.
(\ref{eq:exclusive_scaling2}), 
and from the fact that
the average copy number of the 
dominant specie is 
$[A] \approx g/d$, 
we find that
when the production rate, $g$, is varied while keeping
all other parameters fixed,
$\tau \sim [A]$.
Otherwise,
when the degradation rate, $d$, is varied while all other parameters
are fixed,
$\tau \sim [A]^2$. 
In general,
the switching time is 
$\tau = \tau(k,g,d)$,
while the population size, $[A]$, 
of the dominant specie
depends on both $g$ and $d$.
Thus,
by a suitable variation of the rate constants,
any desired dependence of $\tau$ on
$[A]$
can be obtained.
In particular, 
by increasing $k$,
$\tau$ can be increased with no effect on
$[A]$.
A similar result is obtained when $g$ and $d$ are decreased
by the same factor.
We thus conclude that the population size is only one
of several factors that affect the switching time.
A complete description of the switching time should
include all the relevant rate constants.
 
In Monte Carlo simulations of
a switch system with cooperative
binding, the switching time was found to 
depend exponentialy on the copy number
\cite{Warren2004,Warren2005}.
This is consistent with the discussion above, but requires
a well defined protocol according to which
the rate constants are varied.

\section{The Switch with Bound Repressor Degradation}
\label{sec:BRD}

Consider a different variant of the general switch, 
in which not only free repressors,
but also bound repressors
are affected by degradation.
The bound-repressor degradation (BRD) 
tends to prevent the dead-lock situation in which 
both $A$ and $B$ repressors are bound simultaneously. 
This is due to the fact that degradation removes the bound repressor 
from the system, unlike 
unbinding, where the resulting free repressor may quickly bind again.
It turns out that degradation of bound repressors induces 
bistability not only at the level of the master 
equation but even at the level of rate equations.

\subsection{Rate Equations}

The rate equations that describe the BRD switch
take the form

\begin{eqnarray}
\dot{[A]} &=& g (1-[r_B]) - d [A]-
\alpha_0 [A] \left (1-[r_A] \right) + \alpha_1[r_A]  \nonumber \\
\dot{[B]} &=& g (1-[r_A])-d [B]-
\alpha_0[B]\left(1-[r_B]\right)+\alpha_1[r_B] \nonumber \\
\dot{[r_A]} &=& \alpha_0[A]\left(1-[r_A]\right)-\alpha_1[r_A] 
- d_r [r_A] \nonumber \\
\dot{[r_B]} &=& \alpha_0[B]\left(1-[r_B]\right)-\alpha_1[r_B] 
- d_r [r_B],
\label{eq:brd_switch_rate}
\end{eqnarray}

\noindent
where $d_r$ is the degradation rate of the bound repressors.
Assuming quasi-steady state for the binding-unbinding
processes we obtain
the Michaelis-Menten equations

\begin{eqnarray}
\dot{[A]}&=&{g \over {1+k[B]}} 
- \left(d + {{d_rk}\over{1+k[A]}} \right) [A]  \nonumber \\
\dot{[B]}&=&{g \over {1+k[A]}}
- \left( d + {{d_rk}\over{1+k[B]}} \right) [B],
\label{eq:DBR_switch_michaelis_menten}
\end{eqnarray}   

\noindent
where now
$k=\alpha_0/(\alpha_1+d_r)$. 
Note that the coefficients of $[A]$
and $[B]$
in the second terms in 
Eq.
(\ref{eq:DBR_switch_michaelis_menten})
can be considered as effective degradation rate 
constants.

For steady state conditins,  
Eq.
(\ref{eq:DBR_switch_michaelis_menten})
exhibits the
symmetric solution

\begin{equation}
[A] = [B] = \frac{[ (d+d_rk)^2 + 4dkg]^{1/2} - d - d_rk }{ 2dk}.
\label{eq:symmBRD}
\end{equation}

\noindent
This solution exists for any choice of the parameters.
In addition, in some parameter range, two non-symmetric solutions 
exist. These solutions can be expressed as the solutions
of the quadratic equation

\begin{equation}
dd_r k^2 [A]^2 + (gdk+dd_rk+d_r^2k^2-gd_rk^2)[A]+gd=0.
\label{eq:Asymmetrics}
\end{equation}

\noindent
The condition for the
existence of two different solutions of this equation is

\begin{equation}
(g-d_r)[g(kd_r-d)^2-d_r(kd_r+d)^2]>0.
\label{eq:Asymmetrics2}
\end{equation}

\noindent
In order for them to be positive the 
condition 
$g>d_r$ 
must be satisfied.
Thus, the bifurcation takes place at

\begin{equation}
k_c = \frac{d(\sqrt{g}+\sqrt{d_r})}{d_r(\sqrt{g}-\sqrt{d_r})},
\label{eq:Asymmetrics3}
\end{equation}

\noindent
and the non-symmetric solutions exist for $k>k_c$. 
Linear stability analysis shows that whenever the non-symmetric 
solutions exist they are stable, while the symmetric solution
is stable only for $k \le k_c$.

The steady state populations of free $A$ and $B$ repressors
vs. $k$, for the BRD switch, are shown in 
Fig. \ref{fig6}.
The results of numerical integration of the rate equations
($\times$) are in perfect agreement with the analytical results
derived above (solid line).
We conclude that the degradation of bound repressors induces
bistability, even at the level of rate equations.
The emergence of bistability can be attributed to the 
fact that the effective degradation rate for the minority
specie in
Eq. (\ref{eq:DBR_switch_michaelis_menten})
is larger than the effective degradation rate 
for the dominant specie. 
This tends to
enhance the difference between the population sizes
and to destabilize the symmetric solution
for $k>k_c$.

\subsection{Master Equation} 

The master equation for the BRD switch 
can be obtained from
Eq. (\ref{eq:general_switch_master})
by adding the term

\begin{eqnarray}
\label{eq:BRD_switch_master}
&& d_r [\delta_{r_A,0} P(N_A,N_B,r_A+1,r_B) 
- \delta_{r_A,1} P(N_A,N_B,r_A,r_B)] + \nonumber\\
&& d_r [\delta_{r_B,0} P(N_A,N_B,r_A,r_B+1) 
- \delta_{r_B,1} P(N_A,N_B,r_A,r_B)].  
\end{eqnarray}

\noindent
For steady state conditions we find that BRD tends to suppress 
the peak near the origin of $P(N_A,N_B)$.
For a suitable range of parameters, two separate peaks
appear, which qualitatively resemble those obtained for the
exclusive switch.
However, unlike the exclusive switch, 
there is a narrow corridor with small 
but non-vanishing probabilities
that connects the two peaks via the origin.
As a result, the switching time for the BRD
switch tend to be somewhat shorter than for the
exclusive switch with the same parameters.
The switching times, $\tau$, vs. the
repression strength, $k$, 
are shown in 
Fig. \ref{fig7}.

We now examine in what range of parameters 
this circuit is indeed a switch
according to the master equation. 
Unlike the rate equation where the condition
for bistability is clear
[Eq. (\ref{eq:Asymmetrics3})], 
in the case of the master equation the
notion of bistability is more subtle.
Thus,
in the analysis below we use the following operational criterion.
First we define the two states of the switch.
The $A$-dominated state is defined 
as the set of all states in which
$N_A>2$ and $N_B=0,1$.
Similarly, the
$B$-dominated state is defined by $N_B>2$ and $N_A=0,1$.
The system is considered as a switch
if, under steady state conditions, the total probability 
to be in either of these states is larger than $0.99$.
This leaves a probability of only 0.01 for all
the intermediate states, 
which the system must visit in order to switch between the
$A$-dominated and the $B$-dominated states.
As a result, the switching rate is low.

We used this criterion in order to find the region in the $(k,d_r)$ plane
of the parameter space in which the BRD circuit exhibits bistability.
It was found that the BRD switch exhibits bistability 
for large enough values of $k$, 
as long as the value of $d_r$ is not too different from $d$.
If $d_r/d \ll 1$, the process of bound-repressor degradation is 
negligible and cannot eliminate the dead-lock situation.
If $d_r/d \gg 1$, proteins bind and quickly degrade.
As a result, the population of the dominant specie is reduced
and bistability is suppressed.

Within the parameter range in which the system exhibits bistability,
we examined the dependence of the switching time 
$\tau$ of the BRD switch 
on the
parameters 
$g$, $d$, $\alpha_0$ and $d_r$. 
It was found that $\tau$ exhibits linear dependence on
the production rate $g$ and on the repression strength $k$
(here, $k$ was varied by changing $\alpha_0$, keeping $\alpha_1$ 
and $d_r$ fixed).
The dependence of $\tau$ on the degradation rate $d$ was found to be 
approximately $1/d^2$.
Note that as 
$d$ was veried, 
we kept $d_r=d$ in order that the system remains bistable. 
Since $k$ depends on $d_r$, 
it slightly varied as well.

Unlike the exclusive switch,
where we managed to obtain the scaling properties of 
$\tau$ by a simple argument, 
the BRD switch turns out to be more complicated. 
This is due to the fact that
there are several processes that may 
lead to the flipping of the switch,
such as the unbinding or the degradation of the bound repressor.
A further complication is that the
two repressors can be bound simultaneously.
As a result, we have not managed to obtain
an expression for $\tau$ in the BRD switch.

\section{The Switch with Protein-Protein Interaction}
\label{sec:PPI}

Consider a switch circuit which 
in addition to the mutual repression,
exhibits 
protein-protein interactions (PPI), namely
an A protein and a B protein may form a complex, AB. 
The AB complex is not active as a transcription factor. 

\subsection{Rate Equations}

The PPI switch can be described by the following rate equations

\begin{eqnarray}
\dot{[A]} &=& g (1-[r_B]) - d [A]-
\alpha_0 [A] \left (1-[r_A] \right) + \alpha_1[r_A] -\gamma AB \nonumber \\
\dot{[B]} &=& g (1-[r_A])-d [B]-
\alpha_0[B]\left(1-[r_B]\right)+\alpha_1[r_B] -\gamma AB \nonumber \\
\dot{[r_A]} &=& \alpha_0[A]\left(1-[r_A]\right)-\alpha_1[r_A] \nonumber \\
\dot{[r_B]} &=& \alpha_0[B]\left(1-[r_B]\right)-\alpha_1[r_B].
\label{eq:PPI_switch_rate}
\end{eqnarray}

\noindent
The parameter $\gamma$ is the rate constant for the binding of
a pair of $A$ and $B$ proteins.
The Michaelis-Menten equations take the form

\begin{eqnarray}    
\dot{[A]} &=& \frac{g}{1+k[B]} - d [A] -\gamma [A][B]\nonumber \\
\dot{[B]} &=& \frac{g}{1+k[A]} - d [B] -\gamma [A][B].
\label{eq:PPI_switch_michaelis_menten}
\end{eqnarray}    

\noindent
For steady state conditions,
these equations exhibit a symmetric solution, $[A]=[B]$,
for any choice of the parameters.
It is the solution of 

\begin{equation}
\gamma k [A]^3 + (\gamma+dk)[A]^2 + d[A] - g = 0.
\end{equation}

\noindent
Since all the coefficients of powers of [A] are positive,
this equation has only one positive solution. 
Also, within some range of parameters 
there exist non-symmetric solutions, given by 
the solutions of 

\begin{equation}
d\gamma k[A]^2 + (d\gamma +d^2 k - g\gamma k)[A]+d^2 = 0.
\label{eq:PPnonsym}
\end{equation}

\noindent
The non-symmetric solutions exist only for the range of parameters 
in which 
Eq. (\ref{eq:PPnonsym}) 
has two positive solutions. 
The condition for this can be easily expressed in terms of the coefficients
in 
Eq. (\ref{eq:PPnonsym}). 

As in the case of the BRD switch, 
bistability is observed even
at the level of rate equations. 
Again, the emergence of bistability can be attributed to the 
fact that the effective degradation rate constant for the minority
specie is larger than for the dominant specie, thus
enhancing the difference between the population sizes 
[note that the effective degradation rate constant for 
$A$ is $(d+\gamma [B])$,
while for $B$ it is $(d+\gamma [A])$].

\subsection{Master Equation}

The master equation for the PPI switch 
can be obtained from 
Eq. (\ref{eq:general_switch_master})
by adding the term

\begin{equation}
\gamma [(N_A+1)(N_B+1) P(N_A+1,N_B+1,r_A,r_B) 
- N_A N_B P(N_A,N_B,r_A,r_B)]. 
\label{eq:PPI_switch_master}
\end{equation}

\noindent
For a suitable range of parameters 
the steady state solution of the master 
equation exhibits two separate peaks.
To draw the range of parameters in 
which bistability takes place we apply the
operational criterion used above for the BRD switch. 
We fix 
$g$,
$d$ 
and 
$\alpha_1$ 
and examined the system in the
$(k,\gamma)$ plane. 
The results are plotted in 
Fig. \ref{fig8} (solid line).

For small values of $\gamma$ 
(weak PP interaction), 
the circuit
does not exhibit bistability. 
As the interaction strength increases
the circuit behaves as a switch for a certain range of repression 
strength $k$. 
This range broadens as $\gamma$ is increased.
Unlike the switch systems discussed above,
in which the bistability gets stronger as $k$ is increased,
the PPI switch is bistable
for intermediate values of $k$.
This can be understood as follows.
Recall that the key to the formation of a switch is the
elimination of the dead-lock situation. 
The exclusive and the
BRD switches deal with this situation directly at the bound 
repressor level. 
However the PP interaction does not directly 
affect the bound repressor.
To prevent the possibility of two proteins bound 
simultaneously, one of them should unbind 
and form a complex with a protein of the other specie.
In order for this to happen, the repressors must not
be bound too strongly.
Therefore, the PPI switch works 
at intermediate repression strength. 
As the PPI becomes more effective
(larger $\gamma$) this mechanism applies
at larger values of $k$.

Enhanced switching properties can be obtained by considering
a hybrid system that combines
PPI and exclusive binding.
The resulting switch exhibits bistability
in a broader
range of parameters than the exclusive or PPI switches alone.
The 
master equation for the
exclusive-PPI switch is 
obtained from
Eq. (\ref{eq:exclusive_switch_master})
by adding the term 

\begin{equation}
\gamma [(N_A+1)(N_B+1) P(N_A+1,N_B+1,r_A,r_B) 
- N_A N_B P(N_A,N_B,r_A,r_B)],
\end{equation}

\noindent
which accounts for the PP interaction. 
Numerical results, shown in 
Fig.~\ref{fig8},
indicate that indeed as expected the exclusive-PPI switch is a
better switch than either the PPI or the 
exclusive switch.
The parameter range in which it 
exhibits bistability is broader.
Thus, it is more
robust to variations in the parameters 
than the exclusive or PPI switches.

\section{Cooperative Binding}
\label{sec:cooperative}

Cooperative binding is found in genetic switch
systems such as the phage $\lambda$ switch
\cite{Ptashne1992}.
In this case, transcription regulation is obtained
only when several copies of the repressor are bound
simultaneously.
This situation can be achieved in two ways.
One possibility is that repressors bind to each other and
form a complex, which then binds to the promoter.
The other possibility is that the repressors bind separately,
but those already bound assist the other ones to bind more
effectively.
In the case of cooperative binding, bistability turns out
to appear even at the level of rate equations
\cite{Cherry2000}.

\subsection{Rate Equations}

Switch systems with cooperative binding are commonly 
described by

\begin{eqnarray}    
\dot{[A]} &=& \frac{g}{1+k[B]^n} - d [A] \nonumber \\
\dot{[B]} &=& \frac{g}{1+k[A]^n} - d [B],
\label{eq:general_switch_hilln}
\end{eqnarray}    
            
\noindent
where $n$ is the Hill coefficient.
It corresponds to the number of copies of the transcription
factor which are required in order to perform the repression
process. Here we focus on the case $n=2$, 
and show that these equations exhibit two stable steady state
solutions 
for some range of parameters.
Imposing 
$\dot{[A]}=\dot{[B]}=0$ 
in Eq. (\ref{eq:general_switch_hilln}),
we obtain 

\begin{eqnarray}    
g - d[A] - kd[A][B]^2 &=& 0 \nonumber \\
g - d[B] - kd[B][A]^2 &=& 0.
\label{eq:general_switch_hilln1}
\end{eqnarray}    

\noindent
Subtracting one of these equations from the other we 
find that

\begin{equation}
-d([A]-[B])-kd[A][B]([B]-[A])=0.
\label{eq:AmB}
\end{equation}

\noindent
Looking for a non-symmetric solution for 
which $[A] \neq [B]$,
we divide 
Eq.
(\ref{eq:AmB})
by
$[A]-[B]$.
We 
find that 
$k[A][B]=1$, 
or $[B]=1/k[A]$.
Inserting this into 
Eq. (\ref{eq:general_switch_hilln1}) 
we get an equation for $[A]$:

\begin{equation}
dk[A]^2-gk[A]+d=0.
\end{equation}

\noindent
This equation exhibits two distinct stable solutions 

\begin{equation}
[A]=\frac{gk \pm \sqrt{g^2 k^2-4 d^2 k}}{2dk},
\end{equation}

\noindent
for
$k>4d^2/g^2$.
This means that the system becomes bistable
at the bifurcation point, $k=4d^2/g^2$. 
In addition to these solutions, 
the symmetric solution 
$[A]=[B]$ 
exists for any choice of the parameters.
This
symmetric solution is stable for
$k<4d^2/g^2$ and becomes unstable at the bifurcation point. 

\subsection{Monte Carlo Simulations}

Consider a switch system with cooperative binding with
$n=2$,
in which two proteins of the
same specie bind together to form a complex or dimer.
The repression of $A$ synthesis is done by dimers composed 
of two $B$ proteins and vice versa. 
For example, consider an exclusive switch, in which
the dimers of $A$ and $B$ cannot be bound simultaneously.
To account for stochastic effects, we have studied this
system using Monte Carlo simulations.

The rate constant for the formation of dimers is
denoted by $\gamma_D$. 
It is assumed that dimers cannot dissociate 
into single proteins, but they can degrade.
The degradation rate of
dimers is denoted by $d_D$.
We examined the dependence of the
switching time $\tau$ on all the parameters. 
We found the following properties.
The dependence of $\tau$ on $g$ was found to be 
linear as for the exclusive and BRD switch.
The dependence on $d$ is very weak, 
except for the limit in which $d$ is very large.
This is because the proteins 
tend to form dimers before they have a chance to degrade.
The dependence of $\tau$ on 
$k=\alpha_0/\alpha_1$ 
is found to be well fitted by a quadratic polynomial.
This means that for sufficiently strong repression, 
$\tau \sim k^2$.

The dependence of $\tau$ on the dimerization rate 
$\gamma_D$ 
[Fig. \ref{fig9}(a)]
exhibits interesting behavior. 
For small values of $\gamma_D$ the system 
is not really a switch, because
almost no dimers are formed. 
Therefore the switching time is short. 
For larger values of $\gamma_D$ 
the dimer population increases and 
the system starts to function as a
switch.
The switching time
$\tau$ increases as the switch
becomes more stable. 
But from some point, 
increasing $\gamma_D$ causes $\tau$ to
decrease. 
This is because, very fast dimerization 
helps the minority
specie to form dimers, 
making it more likely to flip the switch.

The dependence of $\tau$ on 
$d_D$ 
[Fig. \ref{fig9}(b)]
was found to be well fitted by a 
cubic polynomial in $1/d_D$.
This means that in the limit of slowly degrading dimers, 
$\tau \sim 1/d_D^3$. 
In the limit of fast dimer-degradation
the system is not bistable, 
because the population of dimers is too small
to make the repression effective.

The switching time for this system was also
studied in Ref.
\cite{Warren2004}, 
where
$\tau$ was presented as a function of 
the average copy number of the dominant specie. 
However, the copy number depends 
in a non-trivial way on the parameters
and cannot be directly controlled.
Therefore, we believe that 
in a systematic study of the switching times,
it is more practical to examine
the dependence of $\tau$ on the parameters themselves.

Note that there is another important realization of cooperative
binding in which the promoter consists of two binding sites.
When a protein binds to one of them it facilitates the binding
of another protein to the second site. 
The effect of this mechanism is qualitatively similar to the 
one shown above for dimers.
In general cooperative binding induces bistability becuase it
forces the minority specie to recruit at least two proteins
in order to flip the switch.
As a result, cooperative binding helps to remove the dead-lock situation
in which both species are suppressed simultaneously.

\section{Response to External Signals}
\label{sec:response}

Until now our discussion considered only 
spontaneous transitions between 
the two states of the switch. 
Here we demonstrate how an
external signal may lead to the flipping of the switch.
In case of the $\lambda$ switch,
such an external signal may be, for example, the 
exposure of {\it E. coli} 
infected by phage $\lambda$ to UV light. 
In the {\it lac} circuit, the external signal indicates 
the presence of lactose.
We assume that the effect of the external 
signal is that one of the proteins
undergoes a conformal change 
that prevents its binding
to the promoter.
When the signal affects the dominant specie,
this may lead to the 
flipping of the switch.
We assume that the conformal change is 
fast and that it lasts for a period
of time determined by the duration of the 
external signal. 

We have performed Monte Carlo simulations, 
where the binding rate $\alpha_0$ of the
dominant specie was set to zero for 
some period of time (the length of external signal).
We calculated the probability for a 
flipping of the switch during $1800$ (s), which is roughly the
time between divisions of {\it E. coli},
as a function of the signal length. 
The results are
shown in Fig.~\ref{fig10}.

For short duration of the signal, the switch 
has a small chance to flip. As
the duration increases the probability to flip 
increases too, and so for a long 
enough signal, the switch will eventually flip as expected 
(the actual switching time depends on 
the parameters of the switch, like
the production rate $g$ or the unbinding rate $\alpha_1$. 
Here we just demonstrated
that in principle the switch will flip states 
in response to an external signal).

\section{Discussion}
\label{sec:discussion}

In the rate equations, the meaning of bistability is clear.
It typically appears as a result of a bifurcation. 
Below the bifurcation there is a single, stable solution, 
which becomes unstable at the bifurcation point, where two
stable solutions emerge.
In case of the toggle switch, one of these solutions is
dominated by $A$ proteins and the other is dominated by $B$ proteins.
Since both solutions are stable, the possibility of spontaneous 
transitions between them due to stochastic fluctuations is not
included in the rate equation model.

The objects that participate in regulatory processes in cells,
namely genes, mRNAs, proteins and promoter sites are discrete
objects, and some of them often appear in low copy numbers. 
This, together with the fact that many of the relevant processes 
such as diffusion, degradation as well as binding and unbinding 
of transcription factors are of stochastic nature, requires to
consider the role of stochastic fluctuations in these regulatory
processes. This can be done by using the master equation or Monte
Carlo simulations. 

In the master equation, bistability is characterized by two 
separate peaks in the probability distribution. These peaks
should be sufficiently far from each other, with low probabilities
in the domain between them. As a result, the flow of probability
between the two peaks is low and the time between spontaneous
switching events is long. In order to qualify as a switch,
the average time between spontaneous switching events must 
be much longer than the time constants of the transcription,
translation and degradation processes in the cell.

For the systems studied here it was found that the general 
switch without cooperative binding does not exhibit bistability
bistability either with the rate equations or with the master
equation.
Two other variants, the BRD and the PPI switch systems,
were found to exhibit bistability both with the rate
equations and with the master equation.
However, the exclusive switch,
which is not bistable at the rate equation level,
was found to exhibit bistability with the master equation.
Thus, in case of the exclusive switch it is clear that
stochastic fluctuations play a crucial role in making
the system bistable. For this system we also found an exact
phenomenological expression for the switching time in terms
of the rate constants of the relevant processes.

Stochastic analysis of genetic networks can be done 
either by direct integration of the master equation
or by Monte Carlo simulations.
The master equation provides the probability distribution
of the population sizes of all the mRNA's and proteins
in the simulated circuit.
It can be considered as a distribution over a large 
number of genetically identical cells. 
The average population sizes and the rates of 
processes are expressed in terms of moments of this
distribution. 
To obtain such distributions from Monte Carlo simulations,
one needs to repeat the simulations a large number of times
and average over them.
This may be inefficient in terms of computer time, and the
statistical errors may be significant.
On the other hand, unlike the master equation,
Monte Carlo simulations enable to follow
the time evolution of a single cell and directly evaluate
quantities such as switching times and oscillation periods.

The number of equations in the master equation set increases
exponentially with the number of proteins and mRNAs 
included in the simulated circuit.
As a result, the master equation becomes infeasible for 
complex networks.
Recently, we have shown that for reaction networks described
by sparse graphs, one can use suitable approximations and
dramatically reduce the number of equations
\cite{Lipshtat2004}.

A related circuit,
the mixed feedback loop, 
in which $A$ is a repressor to $B$
and the $A$ and $B$ proteins bind to form a complex
was recently studied using rate equations
\cite{Francois2004,Francois2005}.
It was found to exhibit bistability 
within a range of parameters.

\section{Summary}
\label{sec:summary}

Genetic switch systems with mutual repression of two
transcription factors, 
have been studied using a combination of
deterministic and stochastic methods.
These system exhibit bistability, namely
two stable states such that 
spontaneous transitions between them are rare. 
Induced transitions take place as a
result of an external stimulus.
We have studied several variants of the genetic
switch, which exhibit
cooperative binding, 
exclusive binding, 
protein-protein interactions 
and degradation of bound repressors.
For each variant we examined the range of parameters
in which bistability takes place.
Numerous studies have concluded 
that cooperative binding is 
a necessary condition for the emergence of bistability
in these systems.
We have shown that a suitable combination of network structure and
stochastic effects gives rise to bistability even
without cooperative binding. 
The average time $\tau$ between spontaneous transitions was evaluated
as a function of the biological parameters.

\newpage
\clearpage


\begin{thebibliography}{10}
\bibitem{Hill}
Hill-function models are simplifications of rate-law equations.
When derived directly from rate laws, 
$n$ is expected to take only
integer values,
which represents the number of transcription
factors required to perform the regulation. 
However, when these models are used for fitting 
empirical data, $n$ is a fitting parameter which may take 
non-integer values.
\bibitem{mRNA}
Extended circuits that include the mRNA level have also been
studied. It was found that 
there is
a perfect agreement between the 
results of the REs for the simplified and
the extended circuits,
and excellent agreement with only a slight difference in peak
shapes in the ME.
\bibitem{TauMEMC}
Results of $\tau$ calculated in this procedure, 
agree with results 
obtained from averaging a large number of Monte Carlo simulations, 
up to a numerical factor of 2. This factor rises from the fact the in
the procedure we described the system can also return from the $B$ dominated
peak to the $A$ dominated peak.
\bibitem{OnlyK2006}
{We assume that the switching time is a function only of $k$, the ratio
  between $\alpha_0$ and $\alpha_1$, and not of $\alpha_0$ and $\alpha_1$
  separately. This is indeed correct in the case of the exclusive switch,
  because we can always choose the time unit as desired and in such a way
  eliminate one of the variables. This assumption may fail in other cases,
  where $k$ is not simply $\alpha_0 / \alpha_1$. For convenience, in this work,
  most of the time $\alpha_1$ was held fixed at a value of $0.01$, $\alpha_0$
  varied, and the results were plotted as a function of $k$.}
\end{thebibliography}

\newpage
\clearpage

\begin{figure}
\caption{
(Color online)
The population sizes of free $A$ and $B$ 
proteins vs. time 
obtained from a Monte Carlo simulation
(a) for the general switch, where
the system exhibits fast transitions between 
its three states; 
(b) for the exclusive switch
The bistable behavior is clearly observed, where the population size 
of the dominant specie is between $20-60$ and the other specie is nearly
diminished. Failed switching attempts are clearly seen.
The typical switching time is in the order of $10^5$ (s) or roughly 1 day.
Bound proteins are also shown. Their fast binding and unbinding events
cannot be resolved on the time scale that is presented.
In both cases,
$g=0.2$,
$d=0.005$,
$\alpha_0=0.2$ 
and
$\alpha_1 = 0.01$ (s$^{-1}$). 
}
\label{fig3}
\end{figure}

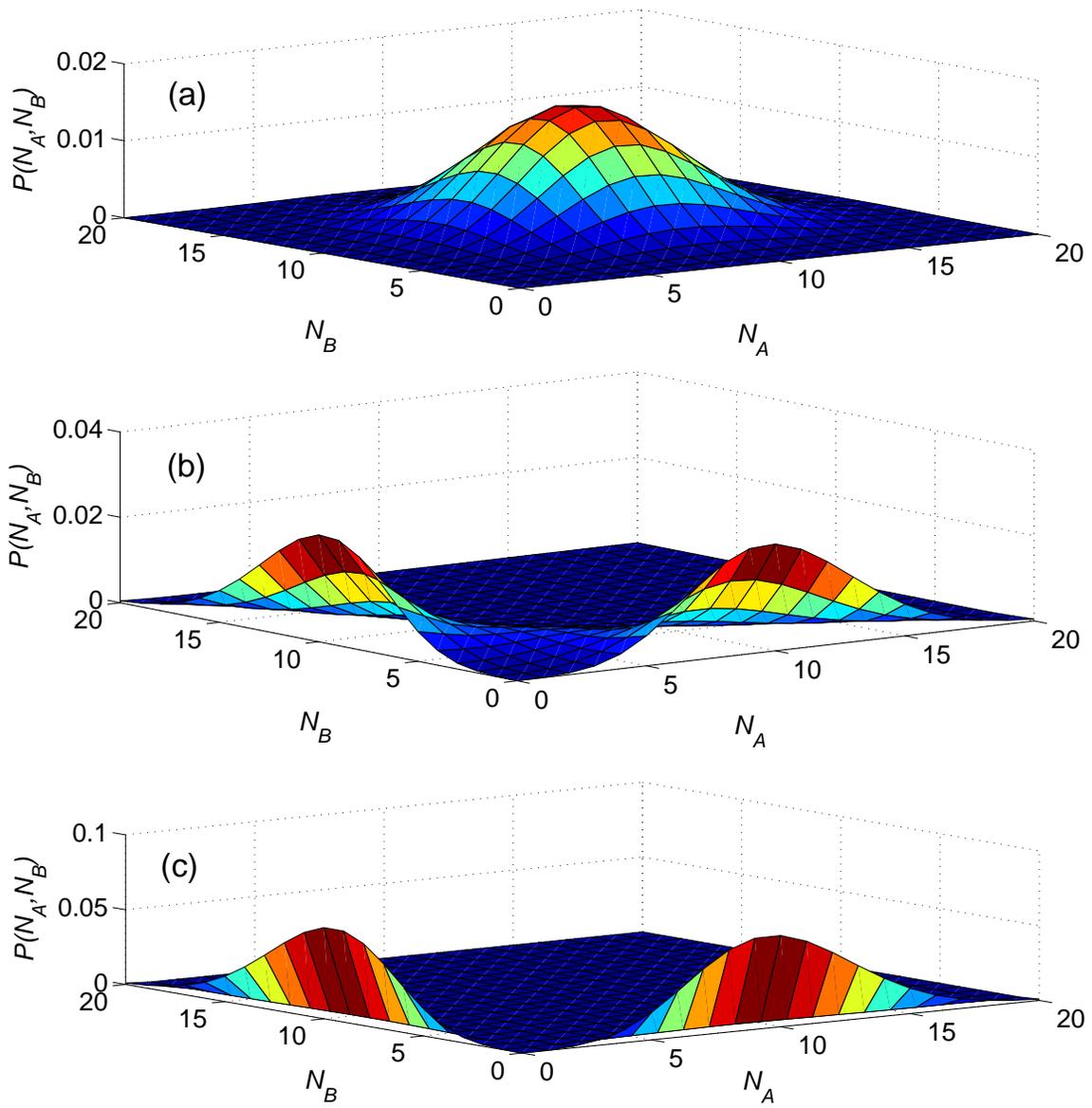
\begin{figure}
\caption{
(Color online)
The probability distribution $P(N_A,N_B)$ 
for the exclusive switch, 
under conditions of 
(a) weak repression ($k=0.005$) where there is one symmetric peak 
(b) intermediate repression ($k=1$) where two distinct peaks begin to emerge 
but are still connected, and
(c) strong repression ($k=50$), where bistability is observed.
}
\label{fig4}
\end{figure}

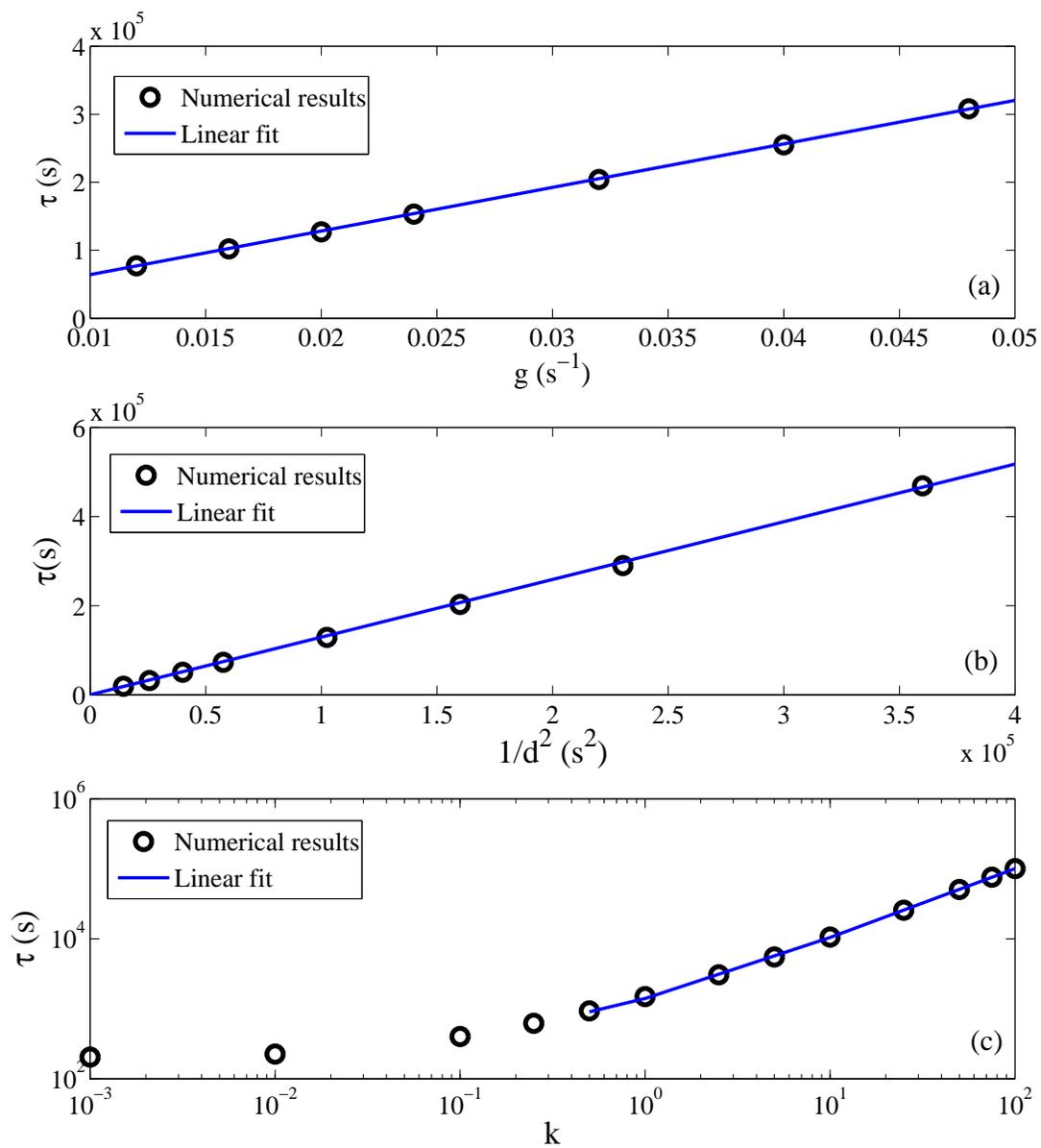
\begin{figure}
\caption{
(Color online)
Scaling properties of the switching time $\tau$
for the exclusive switch
vs. 
the protein synthesis rate $g$,
the degradation rate $d$
and the repression strength $k$.
}
\label{fig5}
\end{figure}

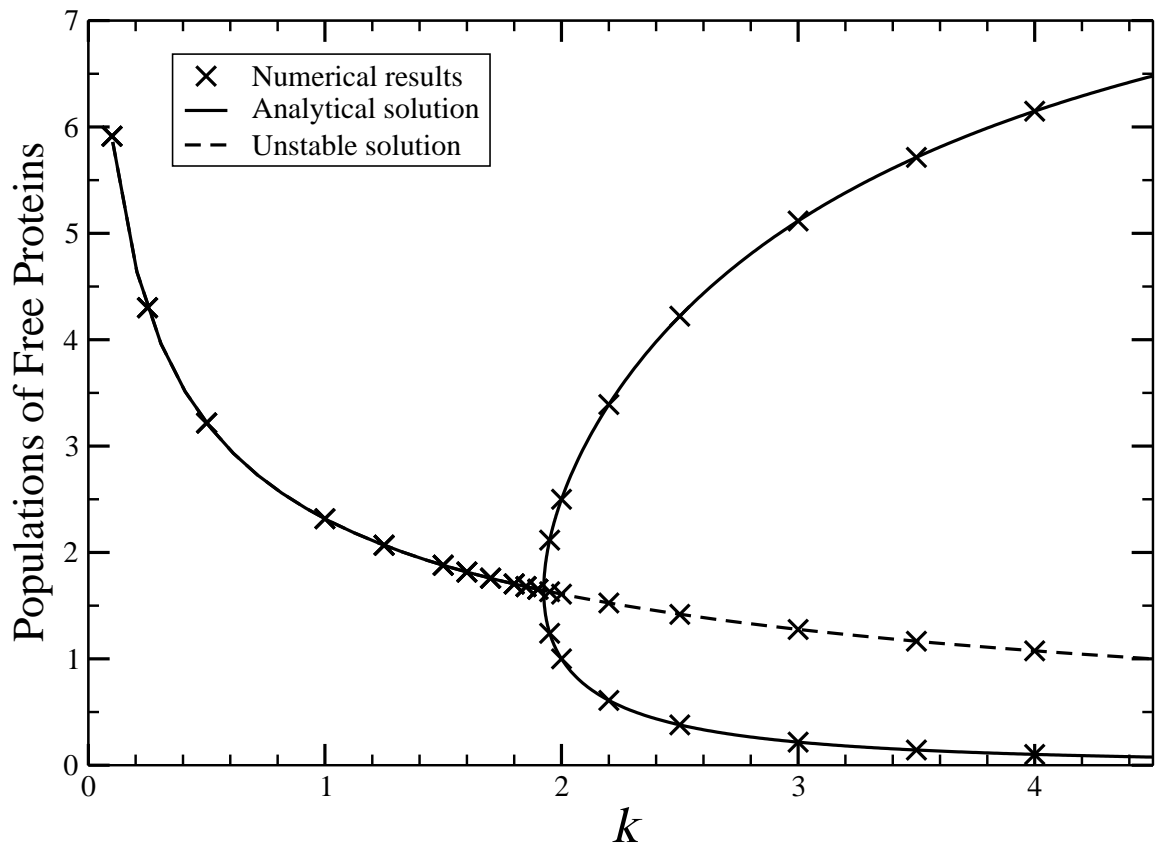
\begin{figure}
\caption{
Population sizes of the free $A$ and $B$ proteins vs. $k$
for the BRD switch obtained from the rate equations.
The parameters are 
$g=0.05$,
$d=d_r=0.005$,
$\alpha_1=0.01$
and
$\alpha_0$ 
is varied.
Here $k_c \approx 1.92$.
Stable solutions are shown by solid lines and unstable solutions
by dashed lines.
}
\label{fig6}
\end{figure}

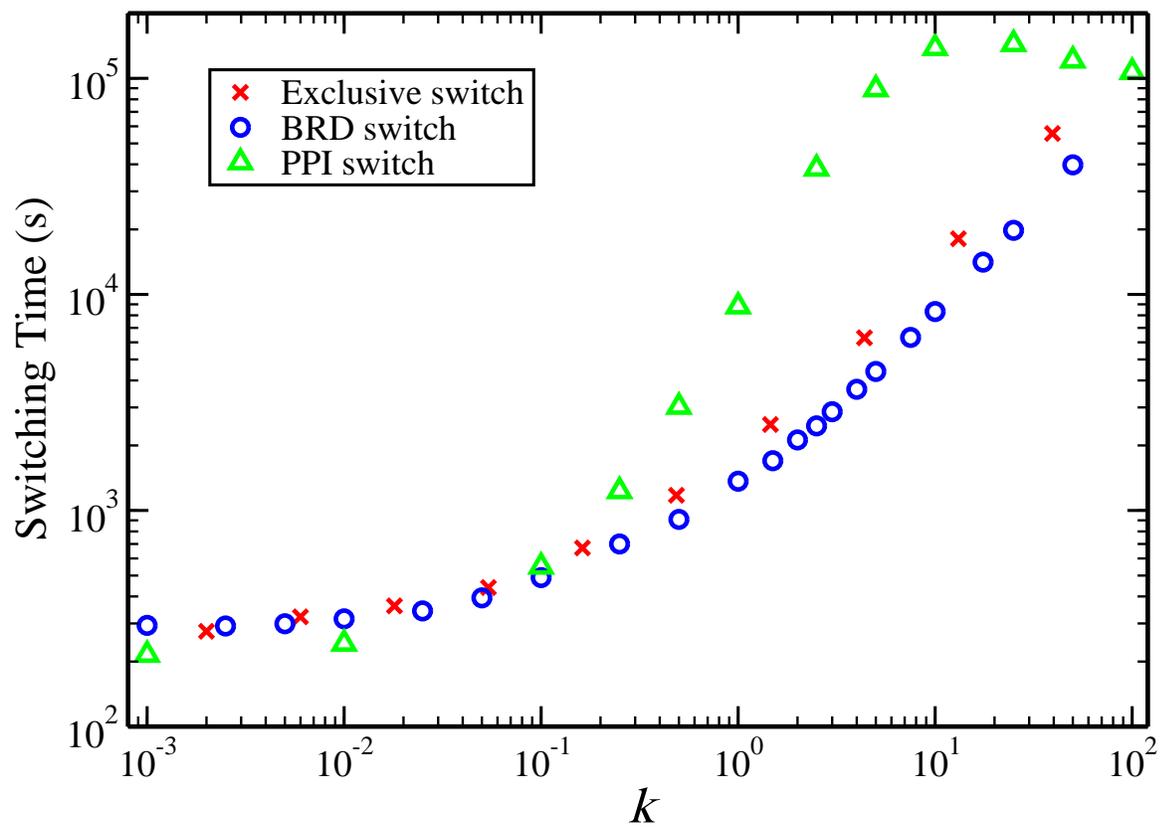
\begin{figure}
\caption{
(Color online)
The switching time $\tau$ vs. $k$ for the 
exclusive ($\times$),
BRD ($\circ$)
and PPI ($\triangle$)
switch systems.
The parameters used are
$g=0.05$, $d=d_r=0.005$ and $\gamma=0.1$ (s$^{-1}$).
}
\label{fig7}
\end{figure}

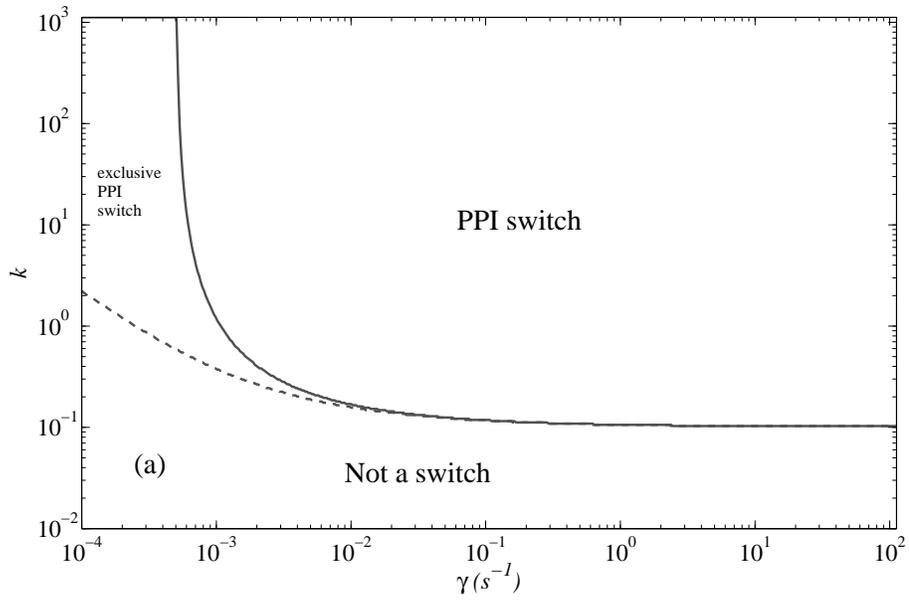
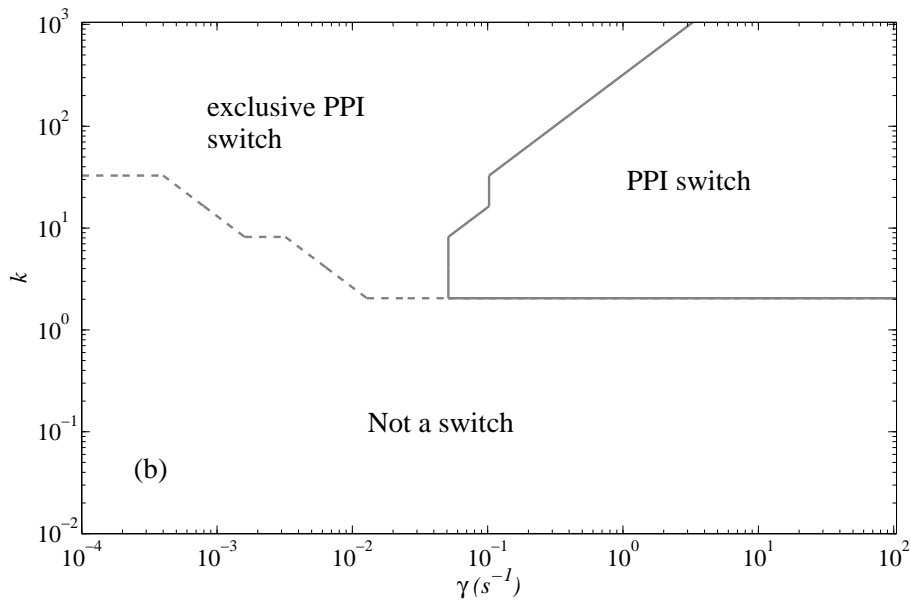
\begin{figure}
\caption{
The range of parameters in the ($\gamma,k$) plane in which bistability
takes place in the PPI switch (solid line) and in the
exclusive-PPI switch (dashed line), using rate equations
(a) and using the master equation (b).
The other parameters are 
$g=0.05$ and $d=0.005$ (s$^{-1}$).
}
\label{fig8}
\end{figure}

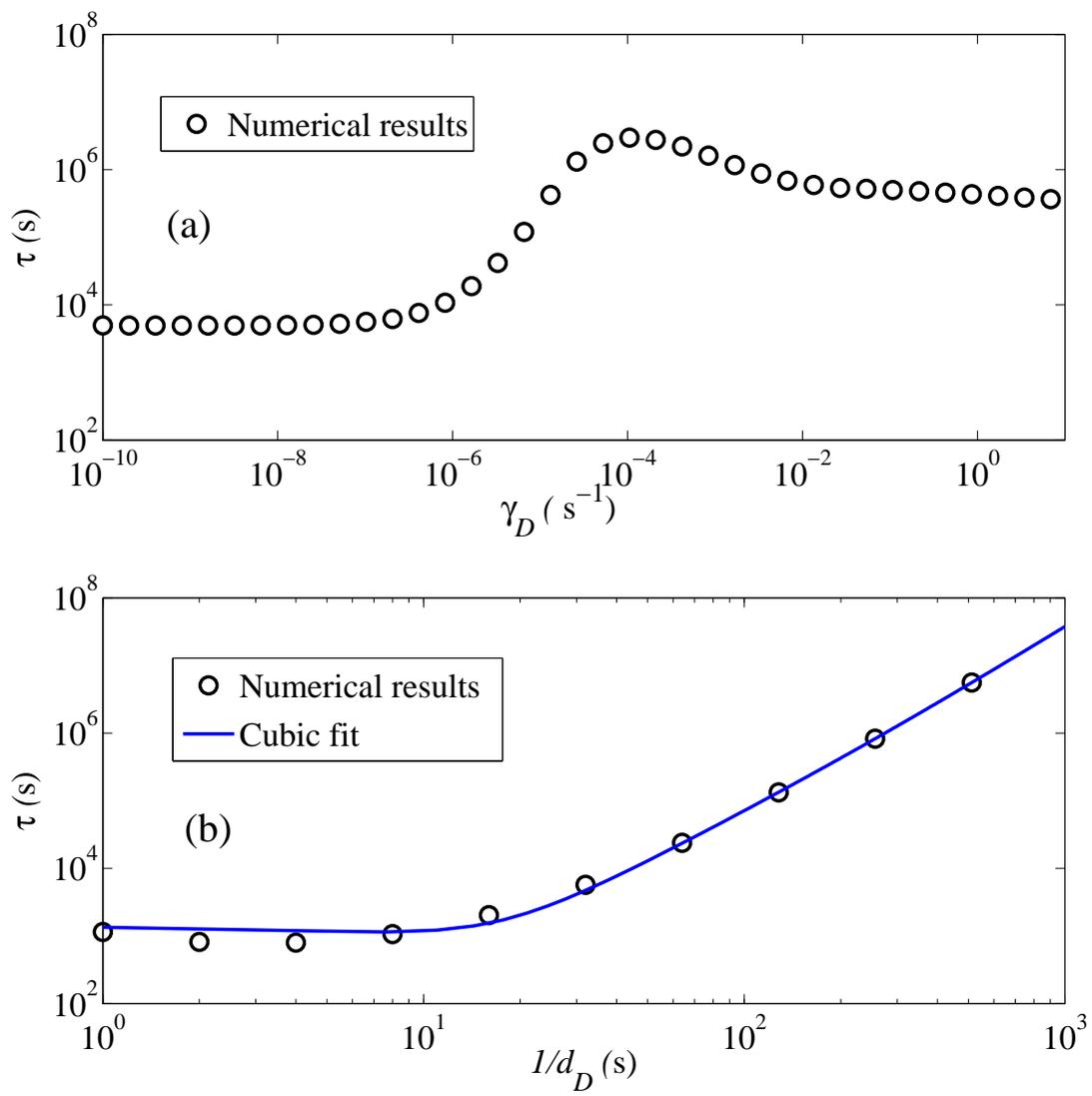
\begin{figure}
\caption{
(Color online)
The dependence of the switching time 
$\tau$ for the dimers exclusive switch
on the 
dimers degradation rate $d_D$ (a)
and the dimerization rate $\gamma_D$
(b).
}
\label{fig9}
\end{figure}

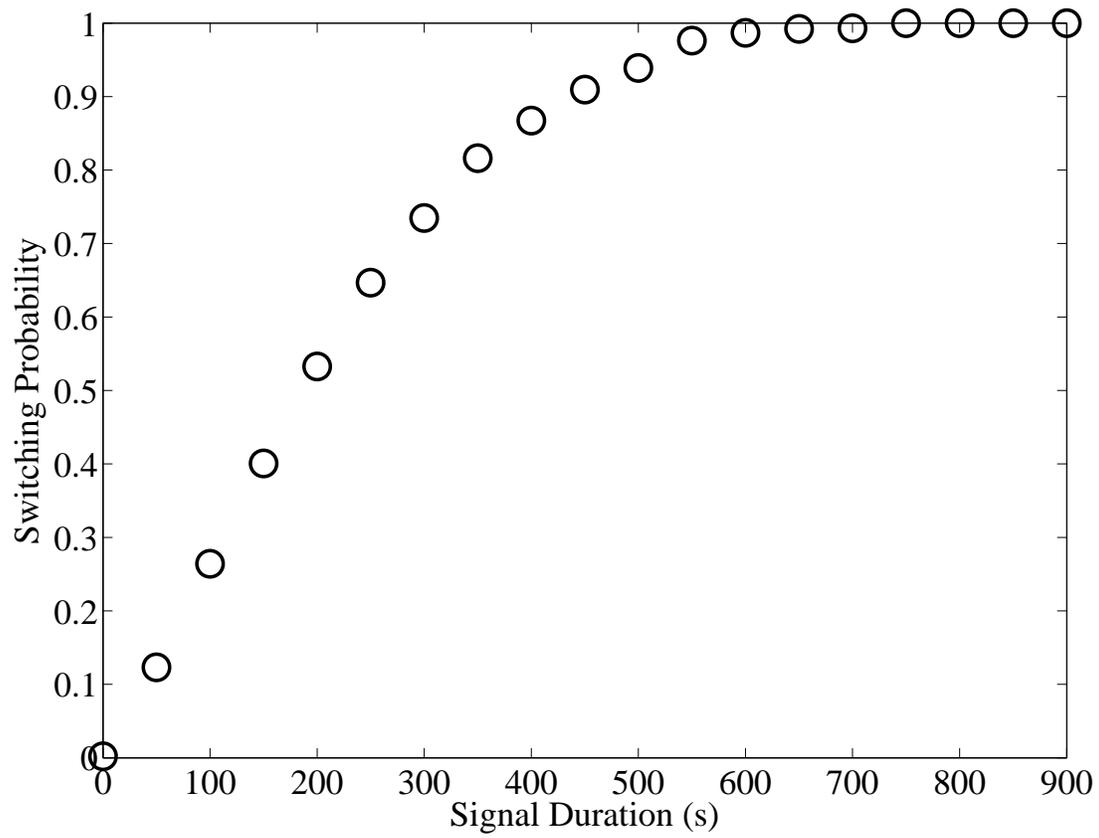
\begin{figure}
\caption{
Probability for the exclusive switch to flip during 1800 (s)
after the initiation of the signal,
as a function of the external signal duration. The parameters used were
$g=0.2$,
$d=0.005$,
$\alpha_1=0.01$ 
and 
$\alpha_0=0.2$ 
or zero during the signal.
}
\label{fig10}
\end{figure}

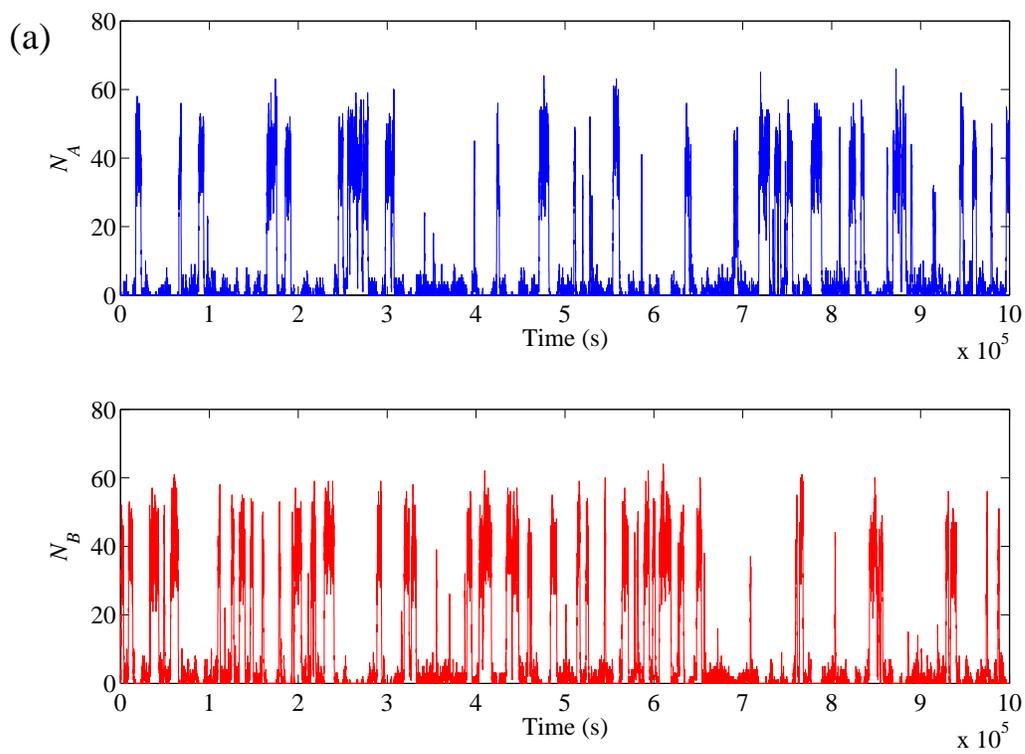
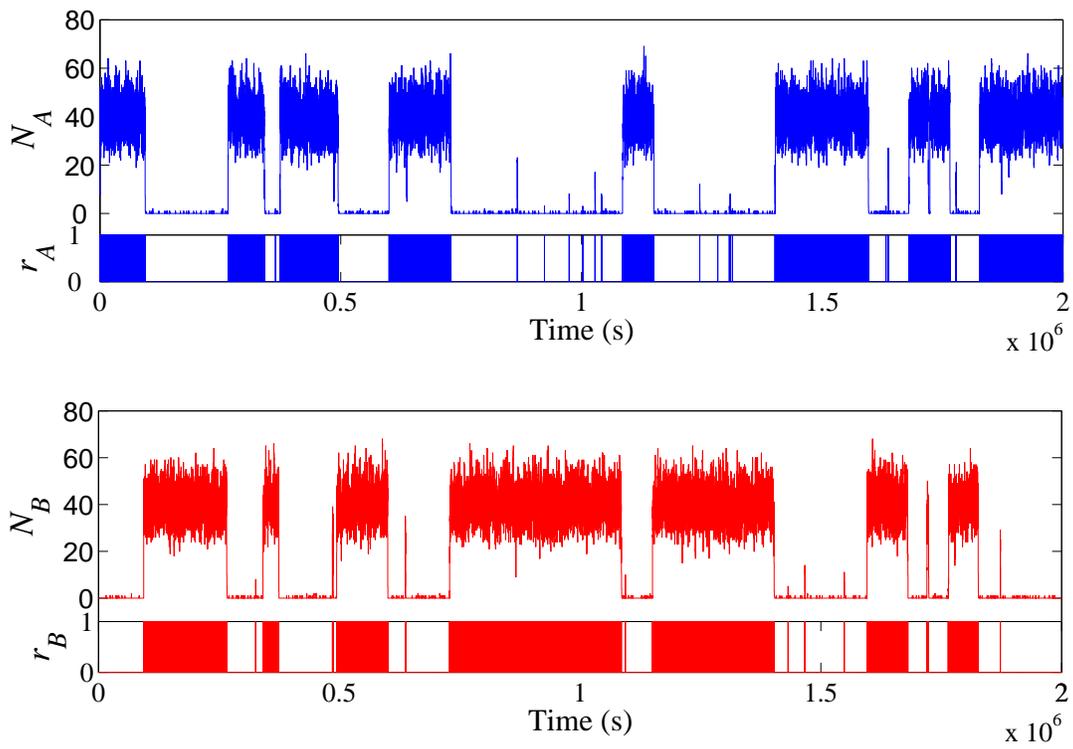
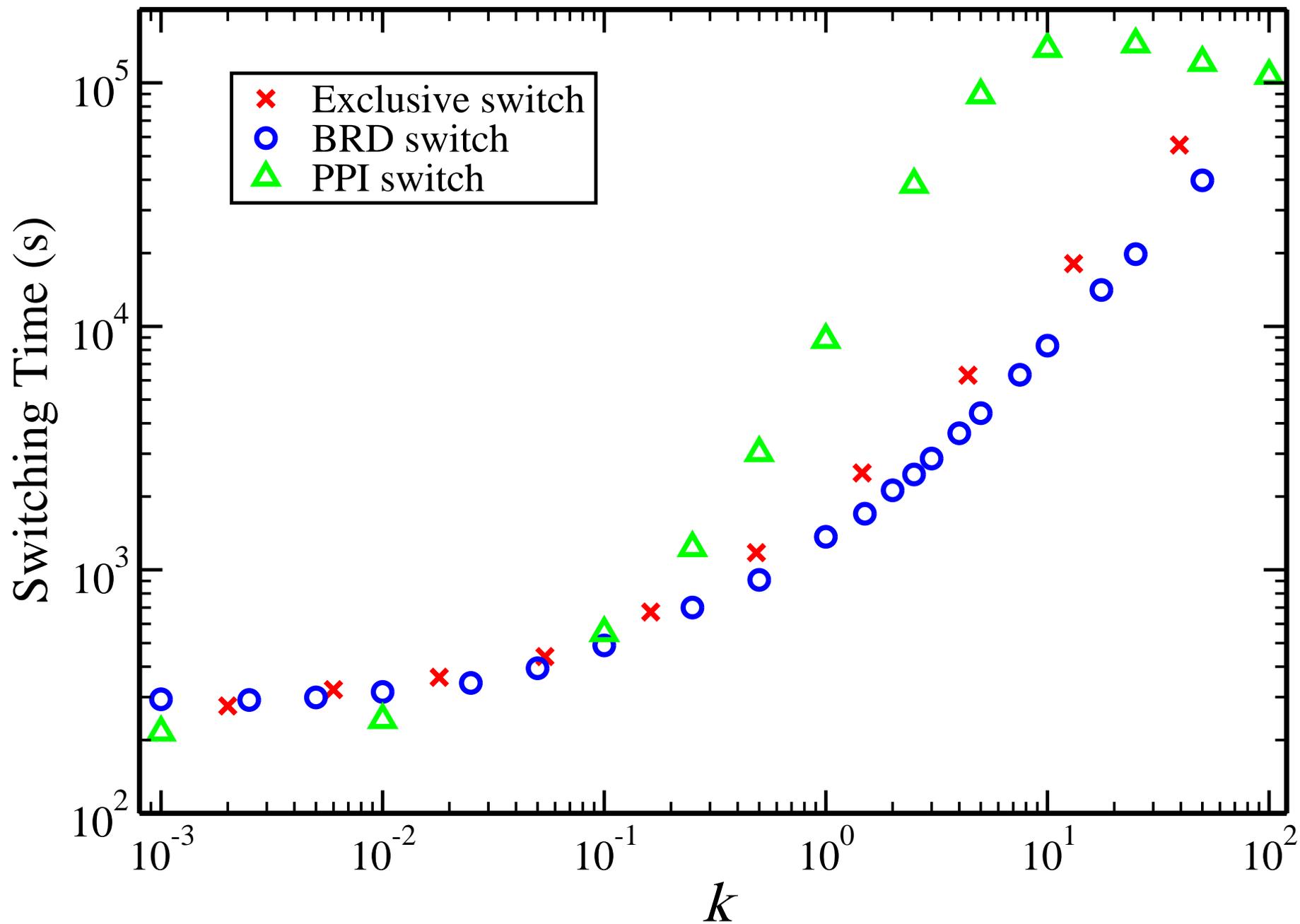
\end{document}